\documentclass[proof]{WileyASNA-v1}

\usepackage{graphicx}%
\usepackage{multirow}%
\usepackage{amsmath,amssymb,amsfonts}%
\usepackage{amsthm}%
\usepackage{mathrsfs}%
\usepackage[title]{appendix}%
\usepackage{xcolor}%
\usepackage{textcomp}%
\usepackage{manyfoot}%
\usepackage{booktabs}%
\usepackage{algorithm}%
\usepackage{algorithmicx}%
\usepackage{algpseudocode}%
\usepackage{listings}%
\usepackage{soul}

\articletype{Article Type}%

\received{17 March 2024}
\revised{26 June 2024}
\accepted{02 July 2024}

\newcommand{\ksmpc}{\,km\,s$^{-1}$\,Mpc$^{-1}$}    
\newcommand{\lcdm}{$\Lambda CDM$} 

\begin{document}

\title{Evaluating a sigmoid dark energy model to explain the Hubble tension}


\author[1]{Torres-Arzayus, Sergio*}

\author[2]{Delgado-Correal, Camilo}

\author[3]{Higuera-G., Mario-A.}

\author[3]{Rueda-Blanco, Sebasti\'an}

\authormark{Torres-Arzayus \textsc{et al}}

\address[1]{\orgname{International Center for Relativistic Astrophysics Network}, \orgaddress{\state{Pescara}, \country{Italy}}}

\address[2]{\orgname{Francisco Jos\'e de Caldas District University of Bogot\'a}, \orgaddress{\state{Bogot\'a}, \country{Colombia}}}

\address[3]{\orgdiv{Observatorio Astron\'omico Nacional}, \orgname{Universidad Nacional de Colombia}, \orgaddress{\state{Bogot\'a}, \country{Colombia}}}

\corres{*Sergio Torres-Arzayus,  \email{sergio.torres@icranet.org}}


\abstract{
In this study we analyze Type Ia supernovae (SNe Ia) data sourced from the Pantheon+ compilation to investigate late-time physics effects influencing the
expansion history, $H(z)$, at redshifts $(z < 2)$. 
Our focus centers on a time-varying dark energy (DE) model 
that introduces a rapid transition in the equation of state, at a specific redshift, $z_a$, from the baseline, $w_{\Lambda} = -1$, value 
to the present value, $w_0$.
The change in the equation of state is implemented as a transition in the DE density scale factor driven by a sigmoid function.
The constraints obtained for the DE sigmoid phenomenological parametrization have broad applicability for dynamic DE models that invoke late-time physics.
Our analysis indicates that the sigmoid model provides a slightly better, though not statistically significant, fit to the SNe Pantheon+ data 
compared to the standard $\Lambda$ cold dark matter (\lcdm) model. 
The fit results, assuming
a flat geometry and maintaining $\Omega_m$ constant at the \emph{2018-Planck} value of $0.3153$, 
are as follows: \(H_0 = 73.3^{+0.2}_{-0.6} \) \ksmpc\ , 
\(w_{0} = -0.95^{+0.15}_{-0.02} \), \(z_a = 0.8 \pm 0.46\). 
The errors represent statistical uncertainties only.
The available SN dataset lacks sufficient statistical power to distinguish 
between the baseline \lcdm\ model and the alternative sigmoid models.
A feature of interest offered by the sigmoid model is that it 
identifies a specific redshift, $z_a = 0.8$, where a potential transition
in the equation of state could have occurred. The sigmoid model does not favor 
a DE in the phantom region ($w_0 < -1$). Further constraints to the dynamic DE model have been 
obtained using CMB data to compute the distance to the last scattering surface. 
While the sigmoid DE model
does not completely resolve the $H_0$ tension, it offers a transition mechanism
that can still play a role alongside other potential solutions. 
}


\maketitle


\section{Introduction}
\label{sec:intro}
\footnote{\emph{Astron. Nachr.}, 2024;e20240034. \\ http://doi.org/10.1002/asna.20240034 }The Hubble tension denotes the disparity between the values of the Hubble constant, 
$H_0$, derived from early universe probes, such as the
cosmic microwave background (CMB) \citep{Bennett_2013} and Baryon Acoustic Oscillations (BAO) data \citep{EISENSTEIN2005360}, 
and those measured using the magnitude-redshift relation with data from
\emph{standard candles} 
(late universe) 
such as Type Ia Supernovae (SNe Ia) 
using the Cepheid distance scale \citep{Riess2022} or calibrated using the tip of the red giant branch (TRGB) \citep{Freedman_2019}.
The first signs of tension between the results from early and late universe probes started showing statistical significance 
when comparing the $H_0$ reported by the \emph{Hubble Space Telescope Key Project} with $H_0 = 74.4 \pm 2.2$ \ksmpc \citep{Freedman_2012}, 
and the Planck mission first release with $H_0 = 67.2 \pm 1.2$ \ksmpc \citep{Planck_2014}, which is a difference of $3\sigma$.
For simplicity, the units of $H_0$ are omitted hereafter (assume \ksmpc).

It is important to clarify that the \emph{Planck} results for $H_0$ are not a direct
measurement but a model-dependent inference, assuming a flat $\Lambda$-cold dark matter (\lcdm) model.
The SH0ES program (Supernovae and $H_0$ for the Equation of State of dark energy)
has presented results of multiple iterations of measurements of $H_0$ using SNe Ia calibrated via
distance-ladder with Cepheids in the hosts of SNe Ia \citep{Riess2022}. Each iteration including a larger sample of SNe Ia and 
improved calibration process. With higher accuracy, the SNe Ia $H_0$ measurements stayed close to the center value (ranging from 73 to 74)
while the errors decreased considerably and the tension growing in significance. 
The Planck + \lcdm result of 2013 for $H_{0}$ yielded a value of $67.3 \pm 1.2$
and later 2015 Planck + \lcdm result, $H_0 = 66.9 \pm 0.6$ \citep{Planck_2016}, and SH0ES, $H_0 = 73.2 \pm 1.7$ \citep{Riess_2016}, 
results yielded a tension at the $3.5\sigma$ level.

Soon after the high statistical significance of the tension was recognized, measurements using independent approaches 
and probes confirmed the tension \citep{Verde_2019}.
More recent results report values
of $H_0 = 73.04 \pm 1.04$ based on SNe Ia \citep[hereafter R22]{Riess2022} and $H_0 = 67.36 \pm 0.54$ derived from CMB
\citep[hereafter Planck-2018]{Planck2018}, results in a discrepancy at the $4.9\sigma$ level of significance.
Comparing more recent $H_0$ measurements using SNe Ia with different calibration approaches reveals a sub-tension within 
results based on distance-redshift analysis. 
The distance ladder approach to determine distances to SNe relies on the use 
of Cepheids or TRGB for calibration. The latter approach yields $H_0 = 69.8 \pm 0.8 (stat) \pm 2.4 (sys)$,
which brings it closer (by $\sim 1.2\sigma$) to the Planck-2018 results \citep{Freedman_2019}.

Given the persistence of the Hubble tension over the past decade, attention has focused, in the theoretical front, on 
possible theoretical models that could explain or alleviate the discrepancy. Dynamic dark energy (DE) models
have attracted interest as they provide a mechanism (via negative pressure) to cause an acceleration that changes in time
and they can be incorporated easily in the Friedmann framework. 
At a high level, these models are grouped into early DE or late-time DE depending on the cosmic epoch 
in which they operate. Early DE models focus on modifications to the pre-recombination physics in the \lcdm\ model \citep{Riess_2023}.
Late-time DE models rely on inflation-like scalar fields that became dominant after CMB decoupling \citep{Avsajanishvili_2024, Shah_2021}.
Beyond scalar field models, there is a plethora of theoretical possibilities that have been explored, including
various flavors of modified gravity, and running constants (time-varying gravitational constant, $\Lambda$, etc.)
For a review see
\citet{DiValentino,Bamba2023,Hu_2023,Knox_2020}. 

While local values of $H_0$, such as those presented in R22, are model-independent,
those derived from CMB depend on physics in the early universe $(z > 1000)$. 
High-definition observations with the James Webb Space Telescope (JWST) firmly exclude the possibility that the Hubble tension is due to systematic errors in distance determination using Cepheids and SNe \citep{Riess_2024}.
Consequently, the challenge presented by the Hubble tension lies in finding models that
preserve the CMB results while allowing for a transition to higher $H_0$ values at low redshifts.
This observation motivates the exploration of late-time physics effects,
which introduce deviations from the standard \lcdm\ model and could potentially elucidate
the $H_0$ tension. 
Furthermore, difficulties with modifications of pre-recombination physics
(i.e. ages of oldest astrophysical objects, cosmic chronometers, multiparameter consistency of early-physics models with CMB data, etc. as presented by \citet{Vagnozzi2023})
strongly point to late-time new physics as a solution (or partial solution) to the Hubble tension.
Additional support for the search of late-time physics effects can be found in analyses of SNe data as presented by
\citet{Dainotti_2021} where they find a decreasing trend in $H_0$ with the redshift of the SNe sample.

Dynamic DE models are described by the equation of state associated to the DE component contributing to the energy density of the Universe.
The equation of state is a ratio of a pressure $P$ to a density $\rho$: $P/\rho$, ($c = 1$).
The result of this ratio is an equation of state (EOS) parameter $w$, with $w = 1/3$ for radiation, and $w = 0$ for non-relativistic matter.
In the \lcdm\ model, acceleration is driven by a cosmological constant $\Lambda$ with an equation of state parameter $w_{\Lambda} = -1$.
In contrast to a cosmological constant, the EOS of dynamic DE models varies with time.
To facilitate the evaluation of dynamic DE models their features can be mapped to a phenomenological representation.
The Chevallier-Polarski-Linder (CPL) dynamic DE model \citep{CPL,Linder_2003},
proposes a simple parametrization for the equation of state involving a linear change with 
the cosmological scale factor, $a$:
$w_{DE}(a) = w_0 + w_a(1 - a)$, where $w_0$ represents the 
value of $w_{DE}$ at the present time, and $w_a$ its slope,
specifically: $dw_{DE}/d \ln (1 + z) |_{z = 1} = w_a/2$.
The parameters $w_0$ and $w_a$ can be determined from fits to SNe data as done for instance by \citep{Torres2023}, who showed
that the CPL parametrization suffers from significant parameter degeneracy,
limiting its ability to explain the tension. 
Moreover, the deviations from the standard \lcdm\ model that the CPL parametrization allows extend over a wide range in redshift space, 
restricting the model's capacity to capture changes in the EOS parameter at specific redshifts.

To address these challenges, we focus on a DE model that introduces a change in the
expansion history (relative to the \lcdm\ model) activated at a specific redshift, $z_a$.
Specifically, we investigate a scenario involving a time-varying DE
that models a rapid change in the EOS parameter 
such that at early times the $w_{DE} = -1$ value is recovered, in agreement with CMB results, 
and at late times it tends to an effective constant value $w_0$ (a model parameter).
The advantage of this late DE model lies in its ability to incorporate a transition at a specific redshift,
thus meeting the requirement to preserve early CMB physics. 
It is worth noting that the proposed model serves as a physics-agnostic
phenomenological parametrization
useful for constraining physical models. 
Examples of such models include a scalar field undergoing a phase transition akin to inflation.
In the context of scalar field models, the sign of the kinetic term in the Lagrangian determines 
the asymptotic behavior of the expansion, specifically, a negative sign (phantom models) 
results in a \emph{big rip}, while a positive sign (quintessence models) results in eternal expansion or 
repeated collapse, depending on the spatial curvature. 
Using the CPL nomenclature with EOS parameters $w_0,w_a$ and $w_0CDM$ for models with $w_{DE} = w_0$ (constant),
quintessence models have a value of $w_0$, with $-1 < w_0 < -1/3$, 
and phantom models have $w_0 < -1$. Quintessence models are further divided, according to 
the rate at which the scalar field evolves, into freezing (slower than the Hubble expansion),
and thawing (faster than the Hubble expansion). 
Recent results from the Dark Energy Survey (DES) \citep{DES_2024} looking at SNe Ia data
and the Dark Energy Spectroscopic Instrument (DESI) \citep{DESI_2024} which makes maps of galaxies, quasars and Lyman$-\alpha$ tracers to 
analyze the BAO signal, find results consistent with a cosmological constant while at the same time not 
excluding flat-$w_0CDM$ models with $w_0$ constant but different than $-1$ or with dynamic DE $w_0w_aCDM$ models.
For $w_0CDM$ models most of the results tend to favor quintessence: DES yields $w_0 = -0.8^{+0.14}_{-0.16}$,
DESI reports $w_0 = -0.99^{+0.15}_{-0.13}$ and \citet{brout2022} using the Pantheon+ SNe Ia catalogue obtains $w_0 = -0.9 \pm 0.14$.
On the other hand, DESI combined with CMB favors phantom models, with $w_0 = -1.1^{+0.06}_{-0.05}$.
For dynamic DE models, DES analysis of flat-$w_0w_aCDM$ models marginally prefers a time-varying EOS with parameters
$(w_0w_a) = (-0.36^{+0.36}_{-0.3}, -8.8^{+3.7}_{-4.5})$, and DESI gives $(w_0,w_a) = (-0.55^{+0.39}_{-0.21}, < -1.32)$. 
Dynamic DE models are therefore still good options, not excluded by data. The question then arises as to 
whether the shape of the time-variation of DE is smooth and continuous in time or has experienced a rapid change at 
a specific redshift. The analysis presented here aims at addressing this important question.

\section{The sigmoid DE model}
\label{sec:sigmoid}
The expansion rate of the universe is defined by the Hubble parameter, $H \equiv \dot{a}/a$,
where $a$ is the cosmological scale factor which is related to the redshift due to the expansion of the Universe, $z$, by 
$a = 1/(1 + z)$. The Hubble constant, $H_0$, is the value of $H$ at the present time, $z = 0$.

The analysis presented in R22 relies on a subset of the Pantheon+ dataset, consisting
of low-redshift SN with $z < 0.15$, providing a measurement of the 
local value of $H_0$. In this redshift range, the Hubble law is 
evident in a straightforward plot of magnitude versus $\log cz$, with
the Hubble constant given by the intercept, $a_B$: 
$\log H_0 = 5 + M_B^0/5 - a_B/5$, where $M_B^0$
signifies the fiducial SN Ia luminosity.
R22's analysis involves calibration parameters in addition to the magnitude-redshift relationship.

Given R22's focus on low-redshift, the approximation for distance at these redshifts is appropriate. 
However, when extending the analysis to higher redshifts, an accurate 
formula for distance becomes crucial. 
Hence, for analysis purposes, it is convenient to categorize the redshift space into three regions:
local, $z < 0.15$, as utilized in R22, 
Hubble Flow (HF), $0.15 < z < 2.3$, determined by the depth of the Pantheon+ dataset, 
and high-redshift, $z > 2.3$.

Fitting models to SNe data in the HF and high-redshift regions requires
model-dependent distance computations.
The magnitude, $m_B$, is linked to distance through the equation: 
\begin{equation}
\label{eq:magmodel}
   m_{B} = 25 + M_0 + 5 \log{d_L \left(z\right) } 
\end{equation}
Here, $M_0$ represents the absolute magnitude of Type Ia supernovae, with R22 determining a value 
for the fiducial SN Ia luminosity $M_B^0 = -19.253$. 
The luminosity distance, $d_L$, expressed in Mpc units,
is model dependent, and for a flat spatial geometry, $\Omega_k = 0$, is given by:
\begin{equation}
\label{eq:ldistance}
   d_L(z) = (1+z) \int_{0}^{z}
   \frac{dz'}{H(z')}
\end{equation}
where $H(z)$ is the Hubble parameter, connected to the cosmological model by the first Friedmann equation, which for flat spatial geometry is:
\begin{equation}
    \label{eq:Friedman1}
    H^2 = \frac{8 \pi G}{3} \rho(z),
\end{equation}
with $G$ the gravitational constant, $\rho(z)$ representing the energy density of all the components contributing to the stress-energy tensor (non-relativistic matter, radiation, and DE).
Friedmann's equation can be written making explicit the dependency of the density terms on $z$ for each component as follows:
\begin{equation}
    \label{eq:friedmann}
    H^2(z) = H^2_0 \sum_{j} \left( \frac{\rho_{j,0}}{\rho_c} f_j(z) \right)
\end{equation}
where $j =$ ``r'' for radiation, ``M'' for matter, and ``DE'' for dark energy, the index ``0'' represents the values at the present time ($z = 0$),
$\rho_c$ is the critical density, $\rho_c \equiv 3H^2_0/8 \pi G$.
The scale factors $f_j$ contain the explicit dependence on $z$, and can be obtained from the continuity relations (conservation of energy) and the equation of state for each component.
The continuity equations are:
\begin{equation}
    \label{eq:continuity}
    \dot{\rho}_j + 3H(\rho_j + P_j) = 0
\end{equation}
With $w_j = P_j/\rho_j$, $H = (1/a)da/dt$, and a change of variable from $a$ to $z$ using $a = 1/(1 + z)$, the continuity equations become:
\begin{equation}
    \label{eq:continuity_z}
    \frac{d\rho_{j}}{\rho_{j}} = 3(1 + w_{j}) \frac{dz}{1 + z}
\end{equation}

For $w$ constant 
the equation above can be solved for $\rho_j$, giving the densities as a function of $z$:
\begin{equation}
    \label{eq:factor_j}
    \rho_j = \rho_{0,j} (1 + z)^{3(1 + w_j)} 
\end{equation}
specifically, $f_r = (1 + z)^4$ 
and $f_M = (1 + z)^3$.
For $w_{\Lambda} = -1$ the equation above automatically returns $f_{\Lambda} = 1$ as expected for 
the EOS parameter of the \lcdm\ model.

For the DE component we are allowing the EOS parameter to vary with time, $w = w_{DE}(z)$. 
The continuity equation for this component (Equation \ref{eq:continuity_z}) can be integrated to solve for $\rho_{DE}$,
from which the factor $f_{DE}$ follows:
\begin{equation}
\label{eq:fz_integral}
    f_{DE}(z) = \exp{\left[\int_{0}^{z}
    \frac{3(1 + w_{DE}(z'))}{1 + z'}
    dz'\right]
    }
\end{equation}
The function $f_{DE}(z)$ for the DE scale factor encapsulates the dependence on the dynamic DE model.

To summarize, the equation for $H(z)$, can be written as $H(z) = H_0 E(z)$, with $E(z)$ given by:
\begin{equation}
\label{eq:Ez}
    E(z) = \sqrt{\Omega_{r} (1 + z)^4 +  \Omega_{M} (1 + z)^3 + \Omega_{DE} f_{DE}(z) }
\end{equation}
The $\Omega_{j}$ parameters denote the standard fractional densities ($\rho_{j,0}/\rho_c$) for radiation, $r$, matter, $M$ and 
dark energy, DE.

We build a phenomenological model by imposing constraints on $f_{DE}(z)$ such that $f_{DE} = 1$ at early times, and for 
late times we allow a behavior of the form $f_{DE}(z) = (1 + z)^{3(1 + w_0)}$, 
with $w_0$ the value of the EOS parameter at the present time.
The function $f_{DE}(z)$ transitions between these two regimes at a redshift $z_a$ (a model parameter).
The desired behavior of the DE scale factor at early times $f_{DE} = 1$ is motivated by CMB results, namely
the factor is constrained to match a spatially-flat \lcdm\ model at early times,
consistent with Planck-2018 results. This CMB constraint depends on details of pre-recombination physics, 
the option to leave the value of $f_{DE}$ at early times as a free parameter is problematic because it would break the 
self-consistency among the 6-parameter \lcdm\ fit used by \emph{Planck} to fit the CMB angular spectrum.
Furthermore, given the high accuracy measurements of the peaks in the CMB power spectrum, we take the \emph{Planck}
results as a firm constraint, hence the $f_{DE} = 1$ choice for high redshifts.

To implement the transition in $f_{DE}(z)$ between the early, $f_{DE} = 1$, and late, 
$f_{DE} = (1 + z)^{3(1 + Q)}$ behavior,
we allow the term $Q$ to take the place of a piece-wise function changing between two constant values (i.e. two regimes: early and late)
at $z = z_a$.  
To avoid numerical instability in the optimization code used in the fit,
the change of value in $Q$ needs to be smooth, we use a sigmoid: 
\begin{equation}
\label{eq:sigmoid}
   Q = w_{0} - \frac{(1 + w_{0})}{1 + e^{\left( \frac{z_a - z}{r}\right)}} 
\end{equation}

The equation above describes 
a smooth transition taking place at a redshift centered around $z_a$ with a transition rate $r$
(fixed to $r = 0.125$ to model a rapid transition).
The model parameters $z_a$ and $w_0$ will be determined from fits to the SN data as described in Section~\ref{sec:fits}.
Note that the constraint is imposed on the DE scale factor $f_{DE}$, not on $w_{DE}$. The shape of $w_{DE}$ can be reconstructed by inverting Equation~\ref{eq:fz_integral}.
In summary, the proposed model has a DE scale factor $f_{DE}$ that changes value in a step-like manner 
from $f_{DE} = 1$ for $z > z_a$ to $f_{DE} = (1 + z)^{3(1 + w_0)}$ for $z < z_a$, 
which implies (subject to satisfying the continuity equation) a DE EOS parameter that changes value in a pulse-like manner
from $w_{DE} = -1$ for $z > z_a$ to $w_{DE} = w_0$ for $z < z_a$.

\subsection{The sigmoid DE mechanism}
\label{sec:mechanism}
\begin{figure}[ht]
   \centering
   \includegraphics[width=78mm,height=11pc]{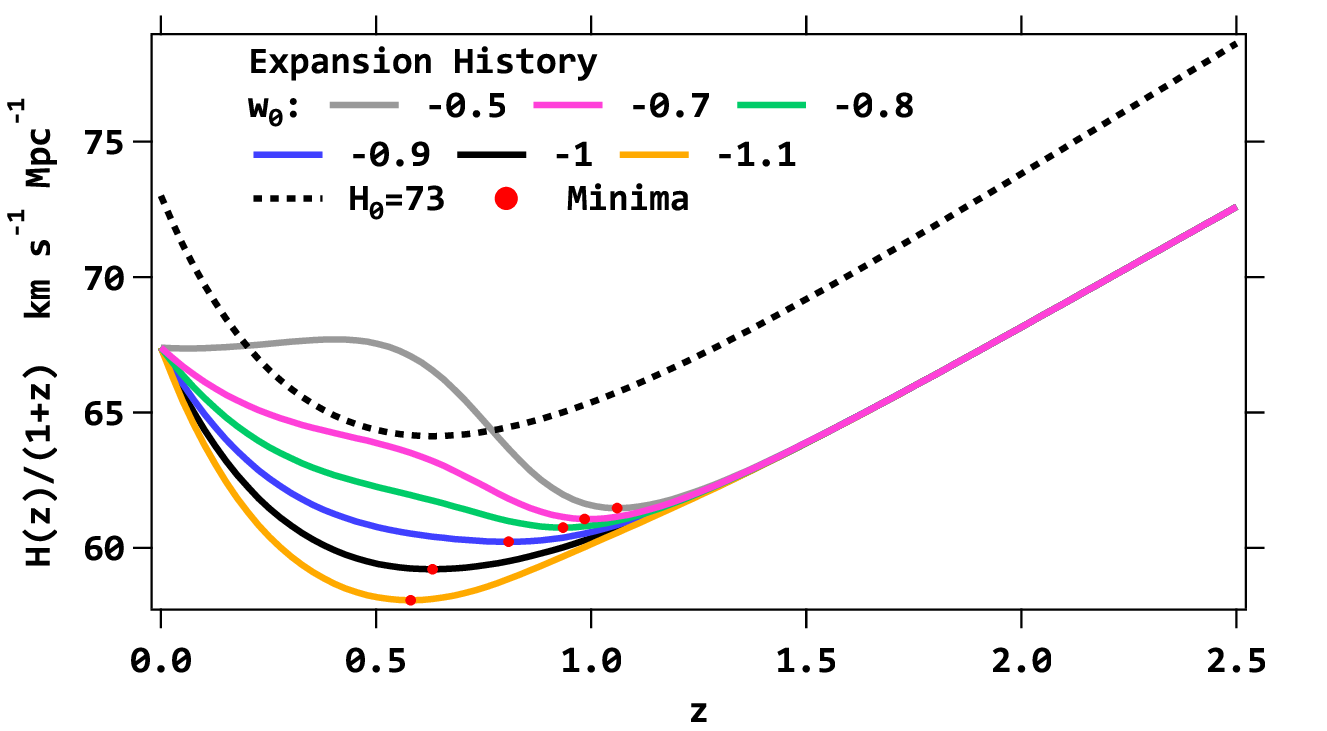}
   \caption{Comoving Hubble parameter as a function of redshift for various settings of the sigmoid parameter $w_{0}$.
   The other parameters of the model were fixed at $H_0 = 67.4$, $\Omega_M = 0.3153$, $\Omega_{\Lambda} = 1 - \Omega_M$, and $z_a = 0.8$.
   The curve with $w_{0} = -1$ (solid black line) represents the \lcdm\ model with $H_0 = 67.4$.
   The black dashed line represents the \lcdm\ model with $H_0 = 73$.}
   \label{fig:history}
\end{figure}

The impact of the sigmoid model on the expansion history, $H(z)$, becomes evident when considering
the shape of $H(z)$ for various settings of the $w_{0}$ parameter.
Figure \ref{fig:history} illustrates curves of $H(z)$ for various settings of the $w_{0}$ parameter. 
The $H_0$ parameter in $H(z) = H_0 E(z)$ acts as the anchor point at $z=0$, 
all the $H(z)$ curves originate from this $H_0$ anchor point and evolve according to the Friedmann framework.
The value of the EOS parameter $w_0$ does not influence $H_0$,
but it can modify the shape of $H(z)$ at intermediate redshifts, as evident from the $H(z)$ curves.
Changing $H_0$ merely shifts the $H(z)$ curves vertically, resulting in
the offset observed between the \lcdm\ model curves (solid and dashed black lines in the figure).
The parameter $H_0$ plays a similar role in the distance equation (Equation~\ref{eq:ldistance}), which 
can be expressed generally as $d_L = A/H_0$, where A is defined by:
\begin{equation}
\label{eq:A_z}
   A(z) = (1+z) \int_{0}^{z} \frac{dz'}{E(z')}
\end{equation}
with $E(z)$ given by Equation~\ref{eq:Ez}. 
When Equation~\ref{eq:A_z} is employed in the fits to compare against SNe data, 
the fitting algorithm's minima tend to bring the ratio $A/H_0$
as close as possible to the data. 
Consequently, to compensate for
an overestimation of the (model-dependent)
term $A$, the fit results in an overestimation of $H_0$.

In the sigmoid DE model, the mechanism functions as follows: 
(i) \emph{If} the true DE behavior of the Universe followed an actual sigmoid
pattern with a parameter $w_0 > -1$, 
the distances (and consequently the magnitudes) of SNe in the HF region
($0.15 < z < 2.3$) would appear smaller than those in a \lcdm\ model.
(ii) When employing a fitting algorithm using a \lcdm\ model with 
SNe observed in our hypothetical true sigmoid universe, the algorithm 
overestimates the term $A$ in distance calculations.
To compensate, the fit pushes $H_0$ to higher values.
As a result, the fit outcomes are biased toward higher $H_0$ values,
partially explaining why the local $H_0$ appears higher than the CMB-derived value.

However, even in the most optimistic scenario where the sigmoid model is accurate, 
we must contend with the fact that the R22 measurement of the local $H_0$ is model-independent. 
The described mechanism could enable the sigmoid model to
explain observations of SNe in the HF redshift region, 
while maintaining a $H_0$ value compatible with the CMB-derived value.
Nevertheless, additional physics, operating in the $z < 0.15$ range, is necessary 
to bring the value of the local $H_0$ closer to the model-independent value measured by R22. 

\section{Fits to Pantheon+ data}
\label{sec:fits}
In the present analysis,
we employed a subset of the Pantheon+ SNe compilation \citep{pantheon}. 
The use of the Pantheon+ offers several advantages: it incorporates 
cross-calibrations of the various photometric systems utilized in the compilation, 
the light curves have undergone a self-consistency analysis process, the 
uncertainties are well characterized with a covariance matrix provided in the data delivery, 
and the data is consolidated in a properly formatted file accessible to the public.
The Pantheon+ dataset has been utilized in recent analyses of the $H_0$ tension, such as R22 and \citep[hereafter B22]{brout2022}.

The Pantheon+ compilation comprises a sample of 1701 light curves from 1550 distinct SNe.
For our study, we selected a subset suitable for analysis in the HF region, implementing a redshift cut
of $0.0233 < z < 1.912$ and conducting several 
data quality checks. The choice of a minimum $z$ value is motivated
by the need to exclude the effects of proper motions and potential local void structures.
A similar $z_{min}$ criterion was applied in R22 when selecting Pantheon+ data for their HF analysis.
Quality checks involved the following criteria: $\sigma_{z} < 0.01$, $|c| < 0.2$, $|x1| < 2.5$, and $\sigma_{m} < 0.5$, where $m$ represents SN magnitude, and 
$c$ and $x1$ denote light-curve fit parameters, for color and shape, respectively. 
After applying these criteria, the resulting HF sample is comprised of 1239 light curves from 1177 distinct SNe.
To fit the sigmoid model to the SNe data, we followed the standard Least-Squares $\chi^2$ 
minimization technique using SN magnitude, $m$,
and redshift, $z$, as the primary data.
The $\chi^2$ value was computed as follows: 
\begin{equation}
\label{eq:chi2}
   \chi^{2} =  \mathbf{R}^{T} \mathbf{C}^{-1} \mathbf{R}
\end{equation}
where $\mathbf{C}$ represents the covariance matrix and $\mathbf{R}$ is the residual vector.
The covariance matrix,
which includes both statistical and systematic uncertainties,
is part of the Pantheon+ data release.  

The residual vector represents the difference between the data, denoted as $m_i$, and the model, denoted as $m_B$:
\begin{equation}
\label{eq:residual}
   R_{i} = m_{i} - m_B(z_i;H_0,w_{0},z_a) 
\end{equation}
where $m_i$ is the apparent magnitude of the $ith$ SNe in the sample, $z_i$ its corresponding redshift.
The model for SN magnitude $m_B$ depends on the sigmoid parameters $(w_{0},z_a)$, and other cosmological parameters
as described in Equation~\ref{eq:magmodel}.

\emph{Planck-2018} established that cosmological parameters are consistent with flat spatial geometry,
and the internal 
consistency of CMB-derived cosmological parameters is tightly constrained
(i.e. it is impossible to alter one parameter without breaking consistency). 
Due to this, $\Omega_M$ is kept constant in the fit and set to the \emph{Planck-2018} value of $0.3153$, 
while the flat geometry assumption is retained, $\Omega_{\Lambda} = 1 - \Omega_M - \Omega_r$.

\begin{figure}[ht]
   \centering
   \includegraphics[width=78mm,height=11pc]{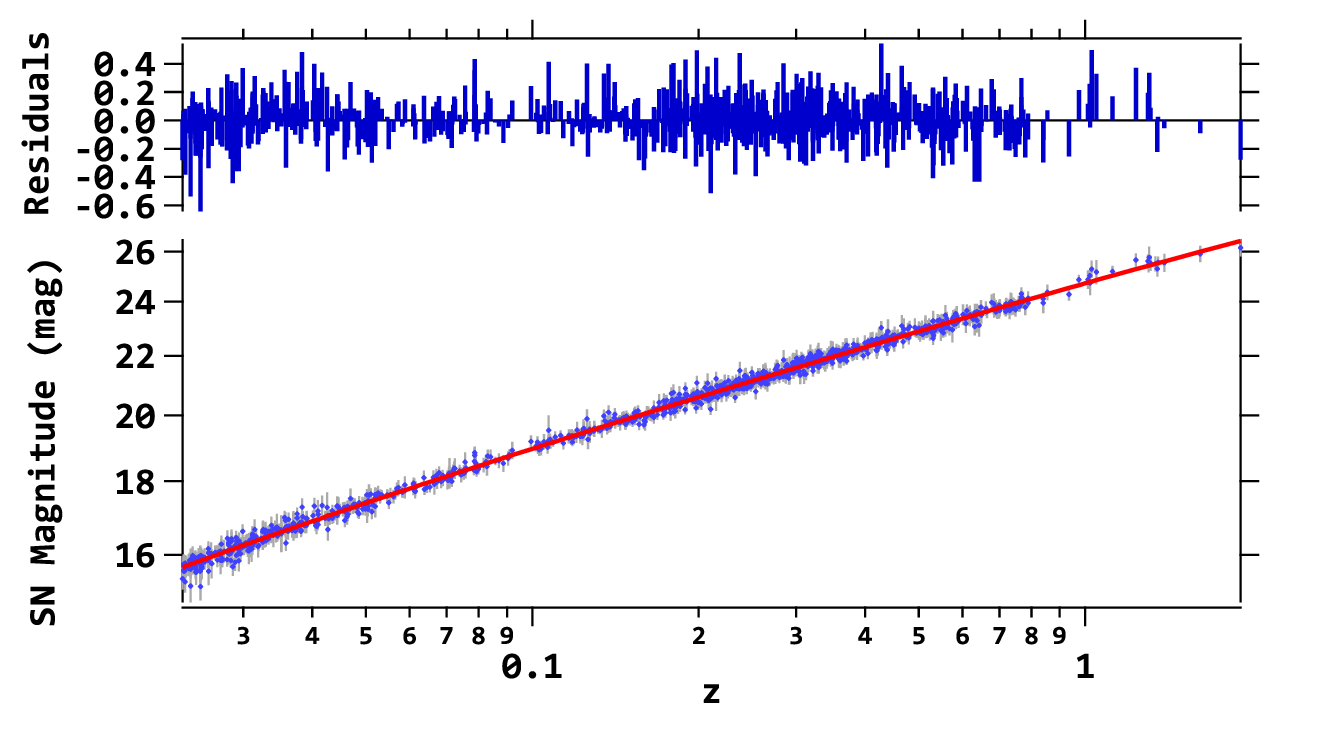}
   \caption{SNe magnitude data (blue dots, and error bars), best fit (red trace) and residuals.}
   \label{fig:fit}
\end{figure}

The fit was performed using a numerical optimization code employing a \emph{line-search} step method and 
\emph{finite-differences} method for calculating the Hessian. The fit parameters 
are $H_0$, $w_{0}$ and $z_a$. 
The uncertainties associated with the fitted parameters were obtained through a Monte Carlo procedure,
as described below, and represent statistical errors only.
The results of the fit are presented in Table \ref{tab:fit} and visually depicted in Figure \ref{fig:fit}.

\begin{table}[h]
   \caption{Fit of the sigmoid and \lcdm\ models to Pantheon+ SNe data. The errors are statistical only.}
   \label{tab:fit}
   \begin{tabular}{lcc}
      \toprule
      Parameter & Sigmoid Fit & \lcdm\ Fit  \\
      \midrule
      $H_0$ & $73.3^{+0.2}_{-0.6}$ & $73.53 \pm 0.15$\\
      $w_{0}$ & $-0.95^{+0.15}_{-0.02}$ & \, \\
      $z_a$ & $0.81 \pm 0.46$ & \, \\
      $\chi^2/N_{dof}$ & $1110.49/1236$ & $1111.42/1238$\\
      \bottomrule
   \end{tabular}
\end{table}

When comparing the fit results presented in Table \ref{tab:fit} it is noteworthy that the sigmoid DE model
yields a slightly lower value of $\chi^2$ relative to the \lcdm\ model. 
However, this difference is statistically inconsequential ($0.26 \sigma$). 
The fact that the $\chi^2$ values in both fits are smaller than their respective numbers of degrees-of-freedom,
$N_{dof}$, renders the $\chi^2$ statistic ineffective as a measure of goodness-of-fit.
In this case, 
it is more likely that the $\chi^2$ values reflect the effects of correlations in the covariance matrix.
In sum, given that for $N_{dof} = 1236$ the $\chi^2$ has a SD of $\approx \sqrt{2 N_{dof}} = 49.7$,
it is highly probable that such small differences ($0.93$) in the $\chi^2$ values are 
the result of noise alone.
Based on these considerations, it is not possible to conclude that the sigmoid model fits 
the data better than the \lcdm\ model. 
However, upon inspecting the residuals, which have an $RMS = 0.143$ mag (smaller than 
the average magnitude error of the sample),
it can be stated that the fit is reasonably good (see Figure~\ref{fig:fit}). Therefore,
the sigmoid DE model is not ruled out; it can explain the data as effectively as the \lcdm\ baseline model.

The sigmoid function identifies a redshift of $z_a = 0.8$ as the time in the 
expansion history when the equation of state transitioned from a cosmological constant, 
$w = -1$, to $w_{0} = -0.95$.
The change in $w$ is small ($0.36\sigma$) and pushes $w$ away from the phantom region ($w < -1$). 

\subsection{Monte Carlo}
\label{sec:montecarlo}
\begin{figure}[ht]
   \centering
   \includegraphics[width=78mm,height=11pc]{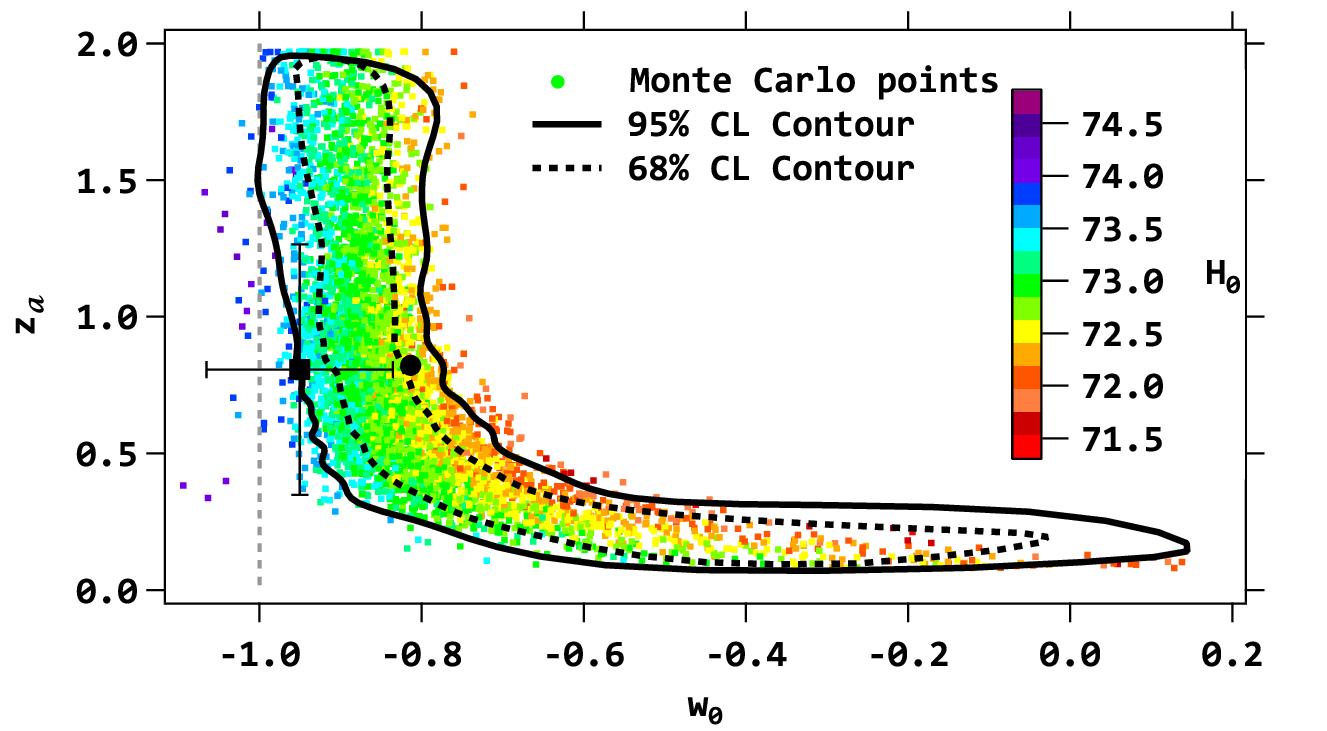}
   \caption{95\% and 68\% confidence level (CL) contours for Monte Carlo points. 
   The color-coded points on the ($w_{0},z_a$) plane represent 
   the result fits to randomized realizations of SN magnitudes at the 
   same redshifts as the sample used in the main analysis. 
   The color scale denotes the corresponding $H_0$ values.
   The black square with error bars represents the best fit sigmoid model, 
   while the round circle is the Monte Carlo average.
   }
   \label{fig:contours}
\end{figure}

\begin{figure}[ht]
   \centering
   \includegraphics[width=78mm,height=11pc]{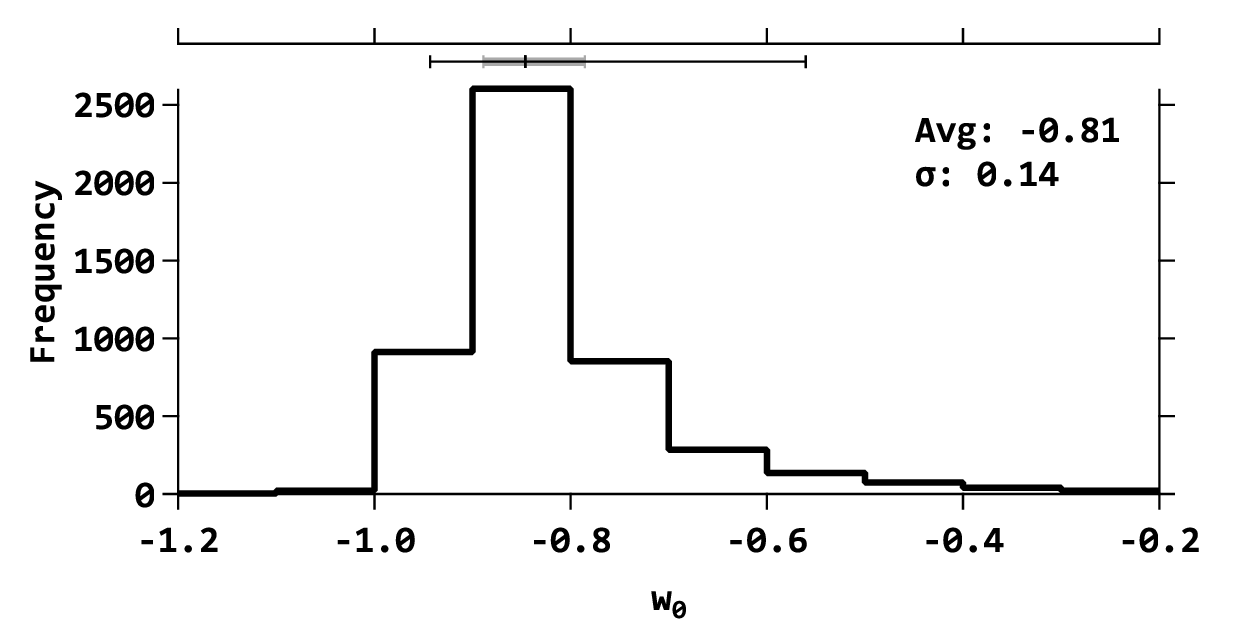}
   \caption{Marginalized distribution of the $w_{0}$ parameter.}
   \label{fig:mcwlate}
\end{figure}

\begin{figure}[ht]
   \centering
   \includegraphics[width=78mm,height=11pc]{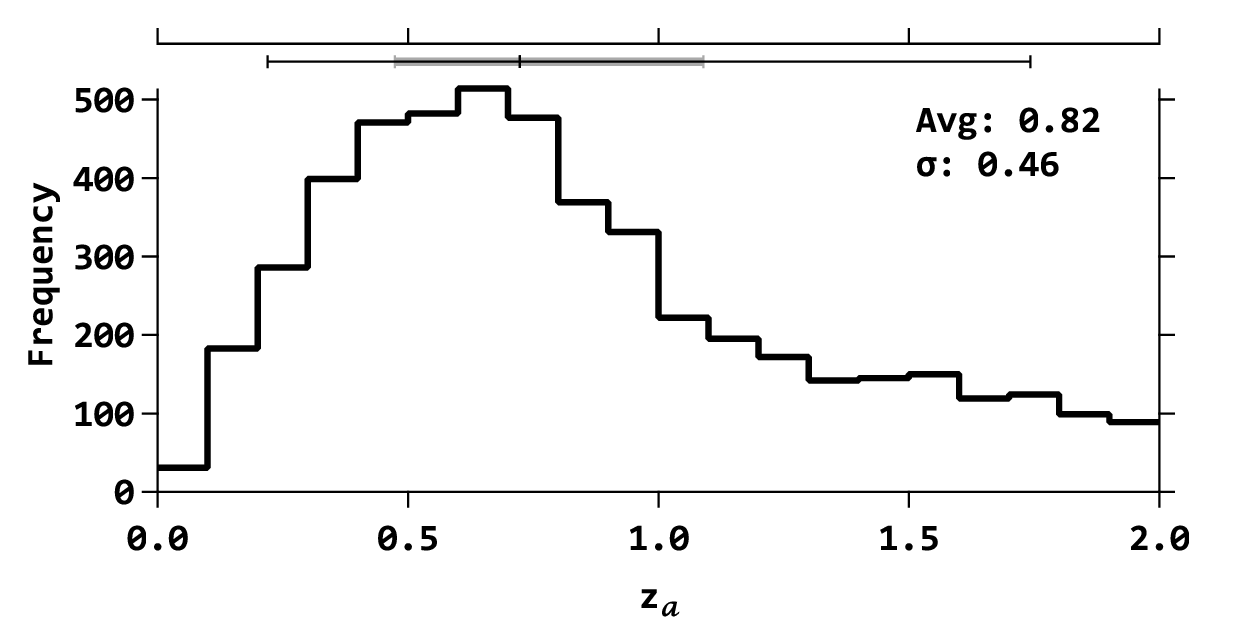}
   \caption{Marginalized distribution of the $z_{a}$ parameter.}
   \label{fig:mcza}
\end{figure}

A Monte Carlo code was developed to generate synthetic SNe compilations
at the same redshifts as the Pantheon+ SNe subset used in the main fit.
For each realization, the Monte Carlo loop generates randomized magnitudes
(as per Equation \ref{eq:magmodel}) with Gaussian noise of $\sigma_m = 0.2$ (the average magnitude error), 
utilizing an underlying DE sigmoid model with true parameters ($w_{0},z_a$)  set equal to the best fit values (Table \ref{tab:fit}).
Subsequently,
the fit code was executed for each realization,
resulting in a corresponding set of best-fit $(w_{0},z_a)$ parameters. 
The 68\% and 95\% confidence level (CL) contours of the Monte Carlo points on the $(w_{0},z_a)$ plane are shown in Figure~\ref{fig:contours}
and the marginalized distributions are presented in Figure~\ref{fig:mcwlate} and Figure~\ref{fig:mcza}.

In Figure~\ref{fig:contours},
the black square with error bars 
represents the best fit, where the error bars 
denote the SDs of the Monte Carlo data (i.e., marginalized errors), and the 
black circle corresponds to the mean of the Monte Carlo points. The relative displacement between these 
two reference points indicates a small bias ($0.037$) in $w_{0}$.
The contour plot illustrates a distinct degeneracy structure in the ($w_0,z_a$) parameter pair. 
The center-line of this structure follows a steep power law. 
For $w_{0}$ values between $-1$ and $-0.8$ the points are 
distributed along a narrow band along the $z_a$ axis, spanning a relatively large range, $0.5 < z_a < 1.9$. 
However, for $w_{0} > -0.8$, the points tend to cluster along a narrow leg parallel to the $w_{0}$ axis,
extending up to $w_{0} \approx 0$. 
The presence of this tail in the distribution causes the aforementioned small bias.
This degeneracy pattern is expected because the effects of DE changes are integrated over redshift space 
(see Equation~\ref{eq:ldistance}). Consequently, if the transition redshift, $z_a$, approaches the present time, 
$z_a \approx 0$, the available range in redshift for late DE (i.e. $w_0$) to operate becomes smaller,
necessitating a larger variation in $w_0$, as illustrated by the horizontal leg on the plot.
Conversely, for high transition redshift ($z_a > 0.8$), $w_0$ is insensitive to the activation redshift, $z_a$, and 
clusters around the $w_0 = -0.95$ band, clearly on the $w_0 > -1$ side, avoiding the phantom region. 
The color scale in the plot corresponds to the values of $H_0$ for each Monte Carlo point. It is observed 
that points with high $H_0 > 73$ (towards the blue end) are grouped toward the $w_{0} < -0.98$ region, 
whereas low $H_0 < 72$ points are clustered
on the opposite side, $w_{0} > -0.8$. 

\subsection{Discussion}
\label{sec:discussion} 
In the optimization code we
computed the $\chi^2$ using the covariance provided in the Pantheon+ release.
We specifically used the
version of the covariance matrix  
(\texttt{Pantheon$+$SH0ES\_STAT$+$SYS.cov})
that encompasses both systematic and statistical errors. 
Entries in the matrix corresponding to SN data removed from 
the sample (as described in Section~\ref{sec:data}) were excluded.
The reported errors for $H_0$ from the fit (as shown in Table~\ref{tab:fit}), $H_0 = 73.3^{+0.2}_{-0.6}$, 
correspond to the $84th$ and $16th$ percentiles of the 
Monte Carlo generated distributions of differences $p_{true} - p_i$, 
representing the difference between the true parameter value and the value on the $ith$ realization.
The SD of the $H_0$ distribution is $\sigma_{H0} = 0.38$ \ksmpc,
which is somewhat lower than the $H_0$ error reported by R22 ($\sigma_{H0} = 1$) and B22 ($\sigma_{H0} = 1.1$). 
This discrepancy in the reported $H_0$ errors is attributed to differences in sample size (due to redshift cuts), and
additional systematics introduced in the R22 results.  
Since our analysis does not incorporate any procedures designed to reduce systematic errors,
we have adopted R22's systematic errors for $H_0$. Consequently, 
our result for the sigmoid model fit is $H_0 = 73.3 \pm 1$ \ksmpc. 

\subsubsection{Comparison with other analyses}
A comparison of the results reported in this study with R22 and B22 
yields valuable insight not only regarding the consistency of the models but
also regarding the SN data's ability to address the $H_0$ tension. Specifically, this comparison sheds light on the 
statistical power inherent in the available SN data for testing models and 
distinguishing between competing alternatives.

Table~\ref{tab:megatable} includes the results of fits where the $\Omega_M$ parameter treated as a free fit parameter, 
as well as the fits reported by R22 and by B22. 
The first observation is that the differences in $H_0$ among these  
fits are not significant (all within $< 0.3 \sigma$).
Secondly, the $\chi^2$ statistics are not discriminative.
It is noteworthy that the $\chi^2$ values are smaller than $N_{dof}$, rendering it an ineffective statistic
for evaluating goodness-of-fit. 
This observation indicates that all the models provide equally good fit 
to the SN data. This conclusion can be restated by asserting that SN data (at least up to a redshift of $\sim 2$,
and given the errors in magnitude, $\sigma_m \sim 0.2$ mag) lack the necessary
discriminatory power to distinguish among competing models. 

\begin{table}
\caption{Summary of fit result statistics.
The "$\Omega_M$" tag indicates that the mass parameter was a free parameter during the fit. R22
does not employ a luminosity distance function parameterized in terms of the standard cosmological model parameters 
(for distance, they use a first order approximation in terms of the acceleration parameter $q_0$).
We calculated the $\chi^2$ associated with B22 (which was not reported in the paper).
$H_0$ represents the best fit value in units of \ksmpc . 
$\chi^2$ denotes the minimum value of this statistic (obtained through the optimization procedure).
$N_{dof}$ is the number of degrees of freedom. 
}
\label{tab:megatable}
\begin{tabular}{lcc}
\toprule
Fit   & $H_0$ &  $\chi^2/N_{dof}$  \\
\midrule
Sigmoid                  & $73.3   \pm 1$    & 1110.49/1236 (0.899)  \\
Sigmoid-$\Omega_M$       & $73.2   \pm 1$    & 1110.45/1235 (0.892)  \\
$\Lambda CDM$            & $73.5   \pm 1$    & 1111.42/1238 (0.898)  \\
$\Lambda CDM$-$\Omega_M$ & $73.3   \pm 1$    & 1110.73/1237 (0.898)  \\
Brout-$\Omega_M$ (B22)   & $73.6   \pm 1.1$  & 1523.02/1699 (0.896)  \\
Riess (R22)              & $73.04  \pm 1.01$ & 3548.35/3445 (1.030)  \\
\bottomrule
\end{tabular}
\end{table}


\section{CMB constraints on the sigmoid DE model}
The CMB angular power spectrum, obtained by the \emph{Planck} satellite, provides precise
estimates of the acoustic angular scale on the sky, denoted as $\theta_*$,
and the comoving sound horizon at recombination, denoted as $r_*$. 
These $\theta_*$ and $r_*$ parameters are determined
by the predecoupling physics of the photon-baryon plasma 
and can impose constraints on cosmological model parameters because they
are linked to the comoving radial distance to the last scattering surface, $d_{LSS}$. 
In flat geometry, 
these parameters are related by a simple geometric construct: $\theta_* = r_*/d_{LSS}$.

To translate CMB measurements of $d_{LSS}$ into constraints on DE model parameters and to assess 
the consistency of the sigmoid model with the CMB, we compare the distance to the LSS predicted 
by the model with the distance obtained from \emph{Planck} data.

The comoving distance is given by
$d_{LSS} = (1 + z_*)d_L(z_*)$, where $d_L(z_*)$ is provided by Equation~\ref{eq:ldistance}.
A baseline value for $d_{LSS}$ is computed using \emph{Planck-2018} values
(from the TT,TE,EE+lowE+lensing result):
$z_* = 1089.92$,
$100\theta_* = 1.04110 \pm 0.00031$, and 
$r_* = 144.43 \pm 0.26$ Mpc,
resulting in $d_{LSS,Planck} = 13872.8 \pm 25$ Mpc. 
This baseline value is then compared with $d_{LSS}$ computed using the best-fit sigmoid parameters (refer to Table~\ref{tab:fit}),
which yields $d_{LSS} = 12741 \pm 153$ Mpc.
This represents a difference of $1132$ Mpc, 
equivalent to $7\sigma$, 
indicating a substantial discrepancy with the baseline.
These results indicate that the best-fit sigmoid model is not consistent with the established cosmological constraints set by CMB physics.

\section{Conclusions}
We explored a potential explanation for the Hubble tension by means of a 
DE model that introduces a deviation in energy density (relative to a pure cosmological constant) at a late-time, low redshift, 
while leaving the expansion history unperturbed for high redshifts.
The proposed model consists of a change in the DE equation of state between two constant values
at a specific redshift, $z_a$. To test the model, we used a subset 
of the Pantheon+ Type Ia supernovae compilation. 
The model's fit to SN magnitude versus redshift data yielded a value for $H_0$ 
of $73.3 \pm 1$, \ksmpc\ and identified a transition redshift of $z_a = 0.8 \pm 0.46$,
where the equation of state parameter $w_{DE}$ transitions from $-1$ to $-0.95$.
Our analysis demonstrates that the available SN data lack the discriminatory power to rule out the \lcdm\ model
in favor of the proposed sigmoid model. 
Despite the test's weak statistical power, the sigmoid model is not
rejected by the data, indicating a potential redshift of interest, namely $z_a = 0.8$, where changes
in the universe's expansion history might have occurred, triggering late-physics effects that could partially explain 
the Hubble tension. 
The fit to the sigmoid model indicates that a late-time deviation of the DE equation of state
(as indicated by the EOS parameter from the fit, $w_0 = -0.95$)
relative to the EOS parameter of the \lcdm\ model ($w_0 = -1$), is significantly limited 
as a candidate to alleviate the Hubble tension. However, the model could still be complementary to other 
late-time physics effects.

\section*{Acknowledgments}
S.R. is the recipient of a scholarship from the 
Observatorio Astron\'omico Nacional, Universidad Nacional de Colombia.

\section*{Data Availability}
\label{sec:data} 

The SNe data used in this study was obtained directly from the Pantheon github site at \\
\texttt{https://github.com/PantheonPlusSH0ES/DataRelease}


\bibliography{h0_bib}


\end{document}